\documentclass[AMA,STIX1COL]{WileyNJD-v2}
\usepackage{dsfont}
\usepackage[utf8]{inputenc}
\usepackage{bm}

\articletype{Commentary/editorial}%

\received{26 April 2016}
\revised{6 June 2016}
\accepted{6 June 2016}

\begin{document}

\title{Conflating marginal and conditional treatment effects: Comments on `Assessing the performance of population adjustment methods for anchored indirect comparisons: A simulation study'}

\author[1]{Antonio Remiro-Az\'ocar}

\author[1,2,3]{Anna Heath}

\author[1]{Gianluca Baio}

\authormark{REMIRO-AZ\'OCAR \textsc{et al}}

\address[1]{\orgdiv{Department of Statistical Science}, \orgname{University College London}, \orgaddress{\state{London}, \country{United Kingdom}}}

\address[2]{\orgdiv{Child Health Evaluative Sciences}, \orgname{The Hospital for Sick Children}, \orgaddress{\state{Toronto}, \country{Canada}}}

\address[3]{\orgdiv{Dalla Lana School of Public Health}, \orgname{University of Toronto}, \orgaddress{\state{Toronto}, \country{Canada}}}

\corres{*Antonio Remiro Az\'ocar, Department of Statistical Science, University College London, London, United Kingdom. \email{antonio.remiro.16@ucl.ac.uk}. Tel: (+44 20) 7679 1872. Fax: (+44 20) 3108 3105}

\presentaddress{Antonio Remiro Az\'ocar, Department of Statistical Science, University College London, Gower Street, London, WC1E 6BT, United Kingdom}

\abstract{In this commentary, we highlight the importance of: (1) carefully considering and clarifying whether a marginal or conditional treatment effect is of interest in a population-adjusted indirect treatment comparison; and (2) developing distinct methodologies for estimating the different measures of effect. The appropriateness of each methodology depends on the preferred target of inference.}

\maketitle

\renewcommand{\thefootnote}{\alph{footnote}}

% \paragraph{Introduction}

The extensive simulation study presented by Phillippo, Dias, Ades and Welton (PDAW)\cite{phillippo2020assessing} is a timely and important addition to the evaluation of population adjustment methods for indirect treatment comparisons, a very relevant and current topic. The authors investigate the statistical performance of matching-adjusted indirect comparison (MAIC), simulated treatment comparison (STC), and a novel method that they have recently proposed\cite{phillippo2020multilevel, phillippo2019calibration} named multilevel network meta-regression (ML-NMR), in an ``anchored'' scenario with a common comparator. The use of these methods in health technology assessment (HTA), both in published literature and in submissions for reimbursement, and their acceptability by national HTA bodies, has increased across diverse therapeutic~areas.\cite{phillippo2019population} 
Their article is an important piece of scholarship because the performance of MAIC, STC and ML-NMR is assessed in a very wide range of settings, which reflect typical scenarios in health technology appraisals and include realistic ``non-ideal'' scenarios under several failures of assumptions. The results of the simulation study are interesting, complement currently available guidance,\cite{phillippo2016nice, phillippo2018methods} and inform the circumstances under which population-adjusted indirect comparisons should be applied.

% Secondly, the statistical performance of ML-NMR is validated in a simulation study and the use of the methodology is~exemplified. 

An important limitation of MAIC and STC is that they are only applicable to pairwise indirect comparisons, and not easily generalizable to larger network structures. This is due to the methods being developed in the context of a two-study scenario, often seen in HTA submissions, where there is one $AB$ study with individual patient data (IPD) and another $AC$ study with only published aggregate data (AgD). In this very sparse network, indirect comparisons are vulnerable to bias induced by effect-modifier imbalances. In larger network structures, multiple pairwise MAICs or STCs do not necessarily generate a consistent set of relative effect estimates for all treatments because the indirect comparisons are made in the AgD study populations, due to the nature of these population adjustment methods. ML-NMR is a timely addition because it extends population adjustment methods to synthesize scenarios where larger networks are required, with the two-study scenario as a special case. The conclusions of the simulation study are that STC and ML-NMR are successful and robust in eliminating bias in the two-study scenario, provided the methods' assumptions hold. On the other hand, MAIC appears to perform poorly in nearly all simulation scenarios. In general, MAIC provides much smaller bias reduction than ML-NMR and STC, and remains biased in relatively favorable settings where we would expect MAIC to perform well.\footnote{In PDAW, MAIC performs poorly even where all effect modifiers are accounted for, there is some covariate overlap between studies (the reference level of the proxy parameter for between-study overlap is set at $\kappa=0.5$, with 50\% of the $AC$ population outside of the $AB$ population), and all other assumptions for population adjustment hold (see Section 4 of PDAW). MAIC remains biased even for the largest overall sample size in the $AB$ and $AC$ trials ($N=1000$ in scenario a), as seen in Table B1 of the online supporting materials, where the induced bias reduces coverage rates below the nominal value. MAIC also produces bias and reduced coverage rates when the covariate-outcome relationship is correctly specified (scenarios e and f), as seen in Table B9 of the supplementary material. For the smallest sample size ($N=100$ in scenario a, see Table B1), and where the correlation between covariates is zero (scenario d, see Table B7), MAIC has substantially greater bias than the standard indirect comparison, even though all effect modifiers have been accounted for.} 

\paragraph{Estimand targeted by the simulation study}

Despite the merits of PDAW, we believe that further clarification of the treatment effect targeted by the simulation study, and of the estimands targeted by the different methods, is required. This is related to the conflation of marginal and conditional (log) odds ratios and the non-collapsibility of this measure of effect. Section 5.3 of PDAW states that the estimands of interest are ``adjusted population-average relative effects between each treatment''. There are two estimands of interest in the comparison of MAIC, STC and ML-NMR. Using the notation introduced by PDAW, these are: $d_{AB(AC)}$, the population-average treatment effect of $B$ vs.\ $A$ in the $AC$ population, and $d_{BC(AC)}$ the population-average treatment effect of $C$ vs.\ $B$ in the $AC$ population, implicitly assumed to be the target population if the shared effect modifier assumption is not made. For ease of exposition, we focus on the former. 

By \textit{population-average} treatment effects, we believe that the authors refer to \textit{marginal} treatment effects. That is, $d_{AB(AC)}$ represents the average difference in the outcome between two identical $AC$ populations, except that in one population all subjects have received treatment $B$, while in the other population all subjects have received treatment $A$, in this case taking the difference on the log odds ratio scale. In health technology assessment, a marginal treatment effect is typically of interest when decision-making is made at the population level. Because the goal is to evaluate the impact of a health technology on the target population for the decision problem, one is interested in the average effect, at the population level, of moving everyone in the target population from treatment $A$ to $B$.

Despite this, the target estimand specified by the data-generating mechanism in Section 5.2 of PDAW is, in fact, calibrated at a different hierarchical level. This estimand is a \textit{conditional} treatment effect, which has a different interpretation than the marginal or population-average treatment effect: a subject-specific interpretation at the individual level. This is because the treatment effects ($\gamma_k$ in Section 5.2 of PDAW) are coefficients of the logistic regression, conditional on the effects of the other covariates that have also been included in the model. That is, $d_{AB(AC)}$ would represent the average effect, at the unit level, of changing an individual's treatment from $B$ to $A$, i.e., the average treatment effect conditioned on the average combination of covariates in $AC$. 
The (log) odds ratio is an effect measure that is not collapsible over covariates. Therefore, the marginal and conditional effects may not coincide, even if there is covariate balance (i.e., independence between the covariates and treatment) and the absence of confounding. This is because the covariate effects do not cancel out in the computation of the conditional log odds ratio, due to the non-linearity of the logit link function of the outcome model. For other link functions, e.g. the identity or log links, the conditional treatment effect does not vary across covariate strata (because the covariate effects cancel out in the computation), and marginal and conditional treatment effects are equal in the absence of confounding. 

Therefore, the article by PDAW assesses the performance of population adjustment methods for anchored indirect comparisons, when the target estimand, $d_{AB(AC)}$, is a \textit{conditional} measure of effect. The article does not evaluate the performance of population adjustment methods, where the target estimand is a \textit{marginal} or \textit{population-average} treatment effect. As seen in the next section, the population adjustment methods considered in the simulation study target different estimand types. This is an important issue that requires further discussion because the appropriateness of each method depends on the target estimand. 

% In our opinion, these issues have misguided the conclusions of the authors.

\paragraph{Methods targeting marginal treatment effects}

MAIC is a method based on propensity score weighting, where the estimate $\hat{d}_{AB(AC)}$ targets a \textit{marginal} log odds ratio, rather than a \textit{conditional} log odds ratio. This is because the outcome model is a univariable weighted regression of outcome on treatment assignment, such that $\hat{d}_{AB(AC)}$ is the fitted treatment coefficient of this weighted regression. This coefficient estimates a relative effect between subjects that have the same distribution of covariates (corresponding to the $AC$ population), assuming a reasonably large sample size and proper randomization in the $AB$ trial, such that (chance) imbalances in baseline characteristics across treatment groups are insignificant. 

Because the simulation study by PDAW targets a conditional treatment effect, the questions raised about the use of MAIC only apply in this context, where the method is biased due to a mismatch in estimands. It is unsurprising that MAIC produces biased results, because it targets a marginal or population-average treatment effect, where the target estimand of the simulation study is a conditional or unit-specific treatment effect. These do not coincide because the measure of effect, the log odds ratio, is non-collapsible.  

If the target of the simulation study is a marginal treatment effect, then we would expect MAIC to be unbiased under no failures of assumptions. In this case, MAIC should also produce unbiased estimates for $d_{BC(AC)}$ because it typically combines marginal treatment effects. Evaluating the performance of MAIC with respect to the other methods, where the target estimand is a marginal or population-average effect, would be informative and impactful. True values of the marginal or population-average effects can be calculated by simulating a very large AgD population (sufficiently large to minimize sampling variability) and fitting a simple univariable regression of outcome on treatment. The treatment coefficient of the regression represents the expected difference in the counterfactual outcomes on the log odds ratio scale if all members of the target population are under active treatment versus if all members of the population are under the common comparator. 

Under correct model specification, MAIC has produced unbiased estimates in recent simulation studies.\cite{remiro2020methods, remiro2020predictive} The authors argue that this is because the studies ``were focused on binary covariates, where issues only arise when covariate proportions are close to zero or one in the IPD study'' and ``were not designed to vary the overlap of continuous covariates between studies, instead basing simulations on scenarios with good overlap''. In our opinion, the difference in the performance of MAIC across simulation studies cannot be attributed solely to differences in overlap. As explained earlier, PDAW consider a sufficiently broad set of scenarios, and MAIC is biased even in relatively favorable conditions with reasonable numbers of ``overlapping'' subjects (effective sample sizes). 

Very recently, after submission and revision of the article by PDAW, we have updated one of the highlighted simulation studies\cite{remiro2020methods} to include smaller sample sizes and to vary the overlap of continuous covariates. The levels of overlap now correspond to percentage reductions in effective sample size that are representative of, and have been inspired by, applications of population adjustment in health technology appraisals. Even though overlap issues can arise more easily with continuous than with binary covariates, MAIC remains unbiased under correct model specification. The study features survival outcomes and a Cox proportional hazards model, which are likely less susceptible to small sample bias than the binary outcomes and logistic regression considered by PDAW. However, we believe that bias is not observed because the target estimand in the study, the indirect effect $d_{BC(AC)}$ (using the notation of PDAW), is a marginal treatment effect. 

\paragraph{Methods targeting conditional treatment effects}

While MAIC uses weighting for population adjustment, STC and ML-NMR are methods based on regression adjustment, which typically targets conditional estimation. The estimate $\hat{d}_{AB(AC)}$ targets a conditional treatment effect, because it is derived from a multivariable regression model containing as predictors both the treatment plus all effect modifiers and, potentially, prognostic variables considered for adjustment. The treatment coefficient of the outcome model is adjusted, controlling for or holding constant all other covariate effects. Therefore, such relative effect has a conditional interpretation. Indeed, ML-NMR is a generalization of the ``gold standard'' correctly specified IPD network meta-regression, which also targets a conditional treatment effect. As a result, it is unsurprising that STC and ML-NMR are adequate estimators of the conditional log odds ratio and produce similar unbiased results where there are no failures in assumptions. Conversely, if the target estimand $d_{AB(AC)}$ was a marginal log odds ratio, we would expect STC and ML-NMR to be biased because these target conditional log odds ratios. 

Unlike ML-NMR, STC has the additional issue of combining incompatible estimates. As commonly implemented, STC combines a conditional treatment effect estimate $\hat{d}_{AB(AC)}$ with a marginal treatment effect estimate $\hat{d}_{AC(AC)}$ in Equation 5a of PDAW,
\begin{equation*}
\hat{d}_{BC(AC)}=\hat{d}_{AC(AC)}-\hat{d}_{AB(AC)}.
\end{equation*}
This is because marginal treatment effects are typically reported in the clinical trial setting, e.g. derived from the coefficient of a simple regression of the clinical outcome of interest on treatment assignment. Clinical trial publications are increasingly reporting conditional or adjusted treatment effects. However, the reported conditional estimate for the $AC$ study is unlikely to be compatible with the conditional effect estimate $\hat{d}_{AB(AC)}$ because these will have been adjusted across different covariate sets and/or using different model specifications. 

Because odds ratios are non-collapsible, conditional odds ratios differ across different sets of adjustment factors and model specifications, even if the covariates and treatment assignment are independent. As mentioned by PDAW, adjusting for a greater number of covariates drives the estimated conditional treatment effect further away from the null, as these account for the some unexplained outcome heterogeneity if they are prognostic of outcome. A conditional estimate from the $AC$ study is only compatible if it is derived fitting the non-centered version of the model in Equation 3 of PDAW,
\begin{equation*}
g(\theta_{ik(AC)}) = \mu_{(AC)} + \bm{x}^\top_{ik(AC)}\boldsymbol{\beta_1} + (\bm{x}^\top_{ik(AC)}\boldsymbol{\beta}_{2,K} + \gamma_C)\mathbb{1}(k=C),
\end{equation*}
including the same explanatory variables in the same manner, to the $AC$ subject-level data. Such data are unavailable and it is unlikely that the estimated treatment coefficient from this model is available in the clinical trial publication. 

ML-NMR is not susceptible to the compatibility issues of STC, with compatible conditional effect measures being combined at each level of the model through the integration in Equation 8d of PDAW. Therefore, ML-NMR is the gold standard for estimating conditional treatment effects, as STC may produce bias if the target estimand is $d_{BC(AC)}$ and incompatible estimates are compared in Equation 5a of PDAW, as observed in a recent simulation study.\cite{remiro2020methods} In that study, the true marginal treatment effect has been set to zero by setting the true unit-level conditional treatment effects for all subjects to zero, implying a ``null''-like simulation setup. As both the true marginal and conditional effects coincide by design, ML-NMR would likely produce unbiased estimates of $d_{BC(AC)}$ in that setup. 

\paragraph{Conclusions}

We believe that this commentary is important because it highlights the importance of: (1) carefully considering and clarifying whether a marginal or conditional treatment effect is of interest in an indirect treatment comparison; and (2) developing distinct methodologies for estimating the different measures of effect. Then, the analyst can decide between each class of methodologies according to the preferred target of inference. Any method is susceptible to bias if targeting the wrong measure of effect. Methods like MAIC are valid for population-based inference, but not ``fit for purpose'' when inference is at the individual level. Similarly, methods like ML-NMR are valid for inference at the individual level, but not designed for population-based inference. 

Note that, throughout this discussion, we have focused on non-collapsible measures of treatment effect, where marginal and conditional estimates may be very different and the conflation of these may lead to bias. Collapsibility does not hold for most measures of interest in population-adjusted indirect comparisons, such as (log) hazard ratios or (log) odds ratios in oncology applications, when investigating time-to-event or binary outcomes, respectively. Nevertheless, clarifying the target of inference is also crucial where the measure of effect is collapsible, in order to ensure adequate uncertainty quantification. 

For instance, in a linear model with continuous outcomes, the treatment effect (mean difference) is collapsible, and the magnitudes of marginal and conditional effects coincide, on expectation, in the absence of confounding. If one runs a simple regression of outcome on treatment assignment, the point estimate of the regression coefficient has both a conditional and a marginal interpretation. However, if one includes further covariates for adjustment, the magnitude of the residuals, residual variance and the standard error of the treatment coefficient are reduced because the covariates, if prognostic of outcome, absorb some outcome heterogeneity that was unexplained in the simple regression. In this case, interpreting the treatment coefficient of the multivariable regression as a marginal treatment effect would lead to an underestimation of uncertainty. In HTA, this means that uncertainty will be incorrectly propagated to the wider health economic model. This is undesirable from the economic modeling point of view and has a negative impact on the ``probabilistic sensitivity analysis'', the (often mandatory)
process used to characterize the impact of the uncertainty in the model inputs on the decision-making process. 

% Probabilistic sensitivity analysis is a required component in the normative
% framework of HTA bodies such as the National Institute for Health and Care Excellence. 

In HTA, when making inferences and policy decisions at the population level, marginal or population-average effect estimates are desirable. Because the policy is to be uniformly applied to the target population, we want the treatment effect to be conditional on the entire population distribution of covariates, such that we marginalize over the individual-level covariates. We are interested in a \textit{single} marginal effect for a specific population, not in a conditional estimate, which is a subgroup-specific measure of effect, allowed to vary across a population at each possible combination of covariates. This may be the case when making treatment decisions, involving solely clinical efficacy or effectiveness, or reimbursement decisions that also take into account cost considerations. For treatment decisions, marginal effects may be of interest if deploying the treatment uniformly across the target population. In reimbursement decisions, marginal effect estimates should inform the mean treatment benefit across the population in a cost-effectiveness analysis. 

Conversely, conditional treatment effect estimates are preferred for inference at the individual level. These are clinically relevant as patient-centered evidence in a physician-patient context, where decision-making relates to the treatment benefit given the specific covariates of an individual patient. Because treatment effects may be heterogeneous across individuals, a finer resolution is of interest in precision or personalized medicine. In the health decision sciences, conditional treatment effects would also be of interest for individual-level microsimulation models that simulate the impact of interventions or policies on individual trajectories. An active area of discussion is whether the primary target of inference in randomized controlled trials should be a marginal or a conditional effect. When approving new treatments, regulators have traditionally focused on marginal estimates as population-level summaries. Nevertheless, conditional effect estimates, adjusted for important prognostic covariates, are increasingly reported as the key analysis. This is due to greater interpretability when applying the results of the trial to individual patients, improved efficiency and statistical power for hypothesis testing, and correcting for (chance) imbalances in baseline characteristics across treatment groups. 

The finding that MAIC has unstable model standard errors and high empirical standard errors under poor overlap between studies or small study sample sizes is an important one. This suggests that the use of MAIC should be avoided when the weights are unstable and dependent on a small number of subjects, i.e., a small effective sample size, a situation that is fairly common in practical applications of MAIC, as evidenced by Phillippo et al. in a recent review.\cite{phillippo2019population} Poor precision and accuracy will be exacerbated in unanchored comparisons, which require accounting for all variables that are prognostic of outcome in order to remove bias. Almost invariably, the level of between-study overlap will decrease as a greater number of covariates are accounted for. MAIC is a weighting method that cannot extrapolate where the between-study overlap is insufficient. However, regression adjustment approaches can extrapolate beyond the covariate space observed in the $AB$ patient-level data. Therefore, the development of a regression adjustment method for indirect comparisons that targets marginal treatment effects would be impactful. 

% A regression adjustment-based estimator inspired by multiple imputation called predictive-adjusted indirect comparison (PAIC) has recently been proposed.\cite{remiro2020predictive} This method extends and adapts the principles of STC so that marginal, instead of conditional, treatment effects are estimated. PAIC splits the adjustment into two separate stages: (1) synthetic datasets are generated, with predicted outcomes drawn from their posterior predictive distribution; and (2) analyses are conducted and pooled on the syntheses by running univariable regressions of predicted outcome on treatment. PAIC has been compared to MAIC in a recent simulation study, yielding more accurate and precise estimates of the marginal treatment effect, particularly under poor overlap and small sample sizes.\cite{remiro2020predictive} PAIC shares many commonalities with ML-NMR. Both methods have been developed under a Bayesian framework and require the same assumptions in the two-study scenario, including the shared effect modifier assumption to generalize the estimated treatment effect to any given target population. PAIC estimates an average treatment effect using simulation and ML-NMR uses numerical integration. Both methods make explicit parametric assumptions about marginal distributional forms and covariate correlations to approximate the joint covariate distribution of the AgD population. 

Currently, to our knowledge, the only peer-reviewed population adjustment methodology that targets marginal effects for indirect treatment comparisons with limited subject-level data is MAIC. MAIC explicitly targets $d_{AB(AC)}$, and care must be taken to ensure that $\hat{d}_{AC(AC)}$ is also a marginal treatment effect, so that compatible estimates are combined in the indirect treatment comparison, and to avoid bias when the measure of effect is non-collapsible. An important weakness of MAIC is that it is only applicable to pairwise indirect comparisons, and it is unclear how to generalize the method for application in larger network structures. Therefore, a research priority is to extend MAIC, or to adapt ML-NMR, so that marginal treatment effects can be estimated in scenarios where larger networks are required. Alternatively, the decision-maker could specify a target population for a specific disease area and clinical outcome, and all manufacturers could conduct their indirect comparisons in the target population. Note that larger networks will be less vulnerable to bias induced by imbalances in effect modifiers, as the heterogeneity ``averages out'' across the network, and the consistency assumption of network meta-analysis can be relaxed using random effects models. 

% When targeting conditional treatment effects, the gold standard population-adjusted estimator for indirect comparisons is ML-NMR. ML-NMR avoids the incompatibility issues of STC in the indirect comparison and can be applied in larger networks of studies and treatments. 

%\nocite{*}% Show all bib entries - both cited and uncited; comment this line to view only cited bib entries;
\bibliography{wileyNJD-AMA}

\end{document}